\documentclass{article}
\usepackage{aip}

\input{tcilatex}

\begin{document}

\author{M.B. Altaie\thanks{%
Electronic Adress: maltaie@yu.edu.jo} \\
%EndAName
Department of Physics, Yarmouk University, Irbid-Jordan}
\title{Back reaction of the neutrino field in an Einstein universe}
\date{December 2002}
\maketitle

\begin{abstract}
The back reaction effect of the neutrino field at finite temperature in the
background of the static Einstein universe is investigated. A relationship
between the temperature of the universe and its radius is found. As in the
previously studied cases of the massless scalar field and the photon field,
this relation exhibit a minimum radius below which no self-consistent
solution for the Einstein field equation can be found. A maximum temperature
marks the transition from a vacuum dominated state to the radiation
dominated state universe. In the light of the results obtained for the
scalar, neutrino and photon fields the role of the back reaction of quantum
fields in controling the value of the cosmological \thinspace constant is
briefly discussed.
\end{abstract}

\section{Introduction}

The discovery of the Cosmic Background radiation (CMB) [1] revived the
theory of the hot origin of the universe (the big-bang model). The CMB was
originally a prediction of the theory of Gamow and his collaborators which
was worked out in the late 1940's in the context of investigating the origin
of the natural abundance of elements. The most refined analysis along this
line predicted a cosmic background radiation at a temperature about $5$ K
(for a concise recent review of the subject see ref. [2]) . However, since
the Gamow model started with the universe at the times when the temperature
was about $10^{12}$ K, the new interest in the origin of the universe sought
much earlier times at much higher temperatures. In the light of the absence
of a full quantum theory of gravity, the works dealing with the state of the
universe at very early times had to consider the quantized matter fields in
the classical background of the universe as described by the theory of
general relativity. Matter fields were brought into connection with
spacetime curvature through the calculation of the vacuum expectation value
of the energy-momentum tensor $<0|T_{\mu \nu }|0>$ \ [3-7]. In these works
and the similar ones that followed, quantum gravity and matter are truncated
at the one loop level. For free matter fields, there are no higher loop
processes anyway. The contribution from graviton is of zeroth order in G. As
the loop expansion is an expansion in \textit{%
%TCIMACRO{\UNICODE{0x127}}%
%BeginExpansion
h\hskip-.2em\llap{\protect\rule[1.1ex]{.325em}{.1ex}}\hskip.2em%
%EndExpansion
, }the theory truncated at the\textit{\ }one loop level contains all terms
of the complete theory of order\textit{\ 
%TCIMACRO{\UNICODE{0x127} }%
%BeginExpansion
h\hskip-.2em\llap{\protect\rule[1.1ex]{.325em}{.1ex}}\hskip.2em%
%EndExpansion
}and is, in that sense, the first order quantum correction to general
relativity\textit{\ }[8]\textit{. }

The motivations for studying the vacuum expectation value of the energy-
momentum tensor stems from the fact that $T_{\mu \nu }$ is a local quantity
that can be defined at a specific spacetime point, contrary to the particle
concept which is global. The energy-momentum tensor also acts as a source of
gravity in the Einstein field equations, therefore $<0|T_{\mu \nu }|0>$
plays an important role in any attempt to model a self-consistent dynamics
involving the classical gravitational field coupled to the quantized matter
fields. So, once $<0|T_{\mu \nu }|0>$ is calculated in a specified
background geometry, we can substitute it on the RHS of the Einstein field
equations and demand self consistency, i.e. 
\begin{equation}
R_{\mu \nu }-\frac{1}{2}g_{\mu \nu }R+g_{\mu \nu }\Lambda =-8\pi <0|T_{\mu
\nu }|0>,  \label{q1}
\end{equation}
where $R_{\mu \nu }$ is the Ricci tensor, $g_{\mu \nu }$ is the metric
tensor, $R$ is the scalar curvature, and $\Lambda $ is the so-called
cosmological constant.

The solution of (\ref{q1}) will determine the development of the spacetime
in presence of the given matter field, for which the vacuum state $|0>$ can
be unambiguously defined. This is known as the ''back reaction problem''. It
is interesting to perform the calculation of $<0|T_{\mu \nu }|0>$ in
Friedman-Robertson-Walker (FRW) models because the real universe is, more or
less, a sophisticated form of the Friedman models. However the
time-dependence of the spacetime metric generally creates unsolvable
fundamental problems. One such a problem was the definition of vacuum in a
time-dependent background [9]; a time-dependent background is eligible for
producing particles continuously, therefore, pure vacuum states in the
Minkowskian sense do not exist. Also an investigation into the
thermodynamics of a time-dependent systems lacks the proper definition of
thermal equilibrium, which is a basic necessity for studying
finite-temperature field theory in curved backgrounds.[10].

The static Einstein universe stands the above two fundamental challenges.
First, being static, the Einstein universe leaves no ambiguity in defining
the vacuum both locally and globally. The same feature also allows for
thermal equilibrium to be defined unambiguously. Furthermore, the Einstein
static metric is conformal to all Robertson-Walker metrics, and this
property enabled Kennedy [10] to show that the thermal Green functions for
the static Einstein universe and the time-dependent Robertson-Walker
universe are conformally related, hence deducing a (1-1) correspondence
between the vacuum and the many particle states of both universes. So that,
under the equilibrium condition, the thermodynamics of quantum fields in an
Einstein universe of radius $a$ is equivalent to that of an instantaneously
static Friedman-Roberson-Walker universe of equal radius [3,6, 11]. This
means that the results obtained in FRW universe would be qualitatively the
same as those obtained in an Einstein universe.

Although not a realistic cosmological model, the Einstein universe remains
to be a useful theoretical tool to achieve better understanding of the
interplay of spacetime curvature and of quantum field theoretic effects. The
recent findings of Plunien \textit{et al}. [12] that finite temperatures can
enhance the pure vacuum effect by several orders of magnitude, can be used
to explain the behavior of the system during the Casimir (vacuum) regime.
Since this means that the finite temperature corrections will surely enhance
the positive vacuum energy density of closed system causing it to behave,
thermodynamically, as being controlled by the vacuum energy.

Recently [13] (hereafter will be referred to as I) we have investigated the
back reaction effects of the conformally coupled massless scalar field and
the photon field at finite temperatures in the background of the Einstein
static universe. In each case we found a relationship between the
temperature of the system and the radius of the Einstein universe. This
relation exhibit a minimum radius below which no self-consistent solution
for the Einstein field equations can be found. A maximum temperature is also
spotted marking the transition from the vacuum dominated state to the
radiation dominated state.

Motivated by the above findings, and in order to complete the picture of the
back reaction of the spinor fields at finite temperatures in the Einstein
universe, we will investigate in this paper the case of the neutrino field
and will deduce the relationship between the temperature and the radius, and
will consequently find the minimum radius and the maximum temperature. Upon
the availability of all results we will compare the contributions of all the
three fields, the scalar, neutrino and the photon fields. Some interesting
results will be discussed.

We will also consider the variation of the cosmological constant with the
temperature of the Einstein universe in presence of the neutrino field, and
then will compile the results for the massless scalar field, the neutrino
field, and the photon field for comparison. The analysis shows that the
smallness of the present value of the cosmological constant is
understandable in the context of the behavior of massless quantum fields in
this universe if they will play the role of energy -momentum source. Also,
we notice that the contribution from the scalar field is higher, by an order
of magnitude, than the contributions of the other fields. Throughout this
paper we will use the natural units in which $G=c=k=$%
%TCIMACRO{\UNICODE{0x127}}%
%BeginExpansion
h\hskip-.2em\llap{\protect\rule[1.1ex]{.325em}{.1ex}}\hskip.2em%
%EndExpansion
\textit{\ }$=1.$

\section{Basic Formalism}

The Einstein static universe is one solution of the modified Einstein field
equations

\begin{equation}
R_{\mu \nu }-\frac{1}{2}g_{\mu \nu }R+g_{\mu \nu }\Lambda =-8\pi T_{\mu \nu }
\end{equation}

\textit{\ }The metric of the Einstein static universe is given by 
\begin{equation}
ds^{2}=dt^{2}-a^{2}\left[ d\chi ^{2}+\sin ^{2}\chi \left( d\theta ^{2}+\sin
^{2}\theta d\phi ^{2}\right) \right] \,,  \label{q2}
\end{equation}
where $a$ is the radius of the spatial part of the universe $S^{3}$ and, $%
0\leq \chi \leq \pi $, $0\leq \theta \leq \pi $, and $0\leq \phi \leq 2\pi $.

We consider an Einstein static universe being filled with massless neutrino
gas in thermal equilibrium at temperature $T$. The total energy density of
the system can be written as

\begin{equation}
<T_{00}>_{tot}=<T_{00}>_{T}+<T_{00}>_{0},  \label{q3}
\end{equation}
where $<T_{00}>_{0}$ is the zero-temperature vacuum energy density and $%
<T_{00}>_{T}\,$is the corrections for finite temperatures, which can be
calculated using the mode sum

\begin{equation}
<T_{00}>_{T}=\frac{1}{V}\sum_{n}\frac{d_{n}\epsilon _{n}}{\exp \beta
\epsilon _{n}+1},  \label{q4}
\end{equation}
where $\epsilon _{n}$ and $d_{n}$ are the eigen energies and degeneracies of
the $n$th state, and $V$=$2\pi ^{2}a^{3}$ is the volume of the spatial
section of the Einstein universe.

The back reaction of the field on the background spacetime can be studied by
substituting for $<T_{\mu \nu }>_{tot}$ on the RHS of the Einstein field
equation (1), i.e.,

\begin{equation}
R_{\mu \nu }-\frac{1}{2}g_{\mu \nu }R+g_{\mu \nu }\Lambda =-8\pi <T_{\mu \nu
}>_{tot}.  \label{q5}
\end{equation}

All the Einstein field equations for the system are satisfied due to the
symmetry of the Einstein universe which is topologically described by $%
T\otimes S^{3}$, and also due to the structure of $<T_{\mu \nu }>$ in this
geometry which is diagonal and traceless, and is given by (see [8],p.186 )

\begin{equation}
<T_{\mu }^{\nu }>=<T_{0}^{0}>diag(1,-1/3,-1/3,-1/3),  \label{q5a}
\end{equation}

Since we are interested in the energy density, we will consider $T_{00}$
only. In order to eliminate $\Lambda $ from (\ref{q5}) we multiply both
sides with $g_{\mu \nu }$ and sum over $\mu $ and $\nu $, then using the
fact that $T_{\mu }^{\mu }=0$ for massless fields, and for the Einstein
universe $R_{00}=0$ , $g_{00}=1$ ,and $R=\frac{6}{a^{2}}$, we get

\begin{equation}
\frac{6}{a^{2}}=32\pi <T_{00}>_{tot}.  \label{q6}
\end{equation}

Note that in the general case conformal anomalies do appear in the
expression for $<T_{\mu }^{\mu }>$ , but because of the high symmetry
enjoyed by the Einstein universe these anomalies do not appear and $<T_{\mu
}^{\mu }>$ is found to be traceless for massless particles.

\section{THE\ VACUUM ENERGY AND BACK REACTION}

The solution of the Dirac equation in an Einstein universe has been
considered by Schrodinger [14] and more recently by Unruh [15]. The eigen
energies are found to be

\begin{equation}
\epsilon _{n}=\frac{1}{a}(n+\frac{1}{2}),
\end{equation}

with degeneracy, for the four-component neutrino

\begin{equation}
d_{n}=4n(n+1),
\end{equation}

The zero-temperature vacuum energy density of the neutrino field in the
background of an Einstein universe is given by [3]

\begin{equation}
<T_{00}>_{0}=\frac{17}{960\pi ^{2}a^{4}}.  \label{q14}
\end{equation}

Neutrinos obey Fermi-Dirac statistics. Thus the total energy density of the
system at finite temperature can be written, in terms of the mode-sum in Eq.
(\ref{q4}), as

\begin{equation}
<T_{00}>_{tot}=\frac{1}{\pi ^{2}a^{4}}\sum_{n=1}^{\infty }\frac{n(n+1/2)(n+1)%
}{\exp (n+1/2)/Ta+1}+\frac{17}{960\pi ^{2}a^{4}}  \label{q15}
\end{equation}

In the low-temperature limit the result reduce to [15]

\begin{equation}
\lim_{\xi \longrightarrow 0}<T_{00}>_{tot}=\frac{17}{960\pi ^{2}a^{4}}.
\label{q16}
\end{equation}

Substituting this in Eq. (\ref{q6}) we get

\begin{equation}
a_{0\nu }=\left( \frac{17}{180\pi }\right) ^{1/2}l_{p}.  \label{q17}
\end{equation}

This is the minimum radius for an Einstein static universe filled with
massless neutrinos at finite temperatures. Note that $a_{0\nu }$\ here is
less than one Planckian length $l_{p}$, this goes beyond the range of
validity of the quasi-classical approximation adopted in the present work.
But fortunately, the region of validity of the approach can be extended if
one takes the number of fields large enough (see for instance Ref. [17]).

In the high temperature (or large radius) limit the result is [16]

\begin{equation}
\lim_{\xi \longrightarrow \infty }<T_{00}>_{tot}=\frac{7}{60}\pi ^{2}T^{4}-%
\frac{T^{2}}{24a^{2}},  \label{q18}
\end{equation}

where $\xi =Ta$.

The back-reaction of the field can be studied if we substitute Eq. (\ref{q15}%
) in Eq. (\ref{q6}), where this time we obtain

\begin{equation}
a^{2}=\frac{16}{3\pi }\sum_{n=1}^{\infty }\frac{n(n+1/2)(n+1)}{\exp
(n+1/2)/Ta+1}+\frac{17}{180\pi }.  \label{q19}
\end{equation}

The solutions for this equation are depicted in Fig. 1, where we see that
the behavior is qualitatively the same as that encountered in the
conformally coupled scalar field case and the photon case in I. Again, two
regimes are recognized: one corresponding to small values of $\xi $ where
the temperature rises sharply reaching a maximum at $T_{max}\approx
1.076T_{p}=1.52\times 10^{32}$ K at a radius $a_{t}\approx
0.204l_{p}=3.2\times 10^{-34}$ cm. Since this regime is controlled by the
vacuum energy (the Casimir energy), therefore we prefer to call it the
''Casimir regime''. The second regime is what we call the ''Planck regime'',
which correspond to large values of $\xi $, and in which the temperature
asymptotically approaches zero for very large values of $a$.

The background (Tolman) temperature of the neutrino field is

\begin{equation}
T_{b\nu }=\left( \frac{45}{28\pi ^{3}a^{2}}\right) ^{1/4}  \label{20}
\end{equation}

At a radius of $1.38\times 10^{28}$ cm (the present Hubble length), we
obtain a background temperature of $23.06$ K, and if we require the
background temperature to have the same value as the average value of the
neutrino background temperature of 1.94 K, the radius of the Einstein
universe has to be $1.94\times 10^{30}$ cm. Again more than two orders of
magnitude larger than the estimated value of Hubble length.

Comparing the temperature for neutrinos from Eq. (\ref{20}) with Eq. (21)
from I, we immediately see that the ratio of the neutrino temperature to the
photon temperature is

\begin{equation}
\frac{T_{b\nu }}{T_{b\gamma }}=\left( \frac{4}{7}\right) ^{1/4}=0.869
\label{21}
\end{equation}

This may be compared with the standard ratio of the neutrino temperature to
the photon temperature, generated in excess because of the heating of the
photons by the electron positron annihilation, calculated according to the
standard big bang scenario (see [18], p.537)

\begin{equation}
\frac{T_{\nu }}{T_{\gamma }}=\left( \frac{4}{11}\right) ^{1/3}=0.714
\label{22}
\end{equation}

Fig 2. compiles the temperature-radius relationships for the scalar,
neutrino and photon fields, and table (1) summarizes the results of the
present work and the previous ones reported in I. Note that we quote here
the correct value for the Tolman temperature of the photon field $T_{b\gamma
}$ which was mistakenly given in I as 30.267 K.

\begin{center}
Table (1) Comparison between the parameters for the massless fields

\begin{tabular}{|l|l|l|l|l|}
\hline
Field & $a_{0}$ & $a(T_{\max })$ & $T_{\max }$ & $T_{b}$ K \\ \hline
Scalar & 0.059 & 0.072 & 2.218 & 31.55 \\ \hline
Neutrino & 0.173 & 0.204 & 1.076 & 23.06 \\ \hline
Photon & 0.279 & 0.340 & 1.015 & 26.53 \\ \hline
\end{tabular}
\end{center}

\section{The Cosmological Constant}

The cosmological constant was introduced by Einstein in order to account for
Mach principle and to justify the equilibrium of a static universe against
its own gravitational attraction. The possibility that the universe may be
expanding led Einstein to abandon the idea of a static universe and, along
with it the cosmological constant. However the Einstein static universe
remained to be of interest to theoreticians since it provided a useful model
to achieve better understanding of the interplay of spacetime curvature and
of quantum field theoretic effects. Recent years have witnessed a resurgence
of interest in the possibility that a positive cosmological constant $%
\Lambda $ may dominate the total energy density in the universe (for recent
reviews see [19] and [20]). This interest stems from the recent observations
of high redshift type Ia supernovae, which appear to suggest that the
universe is accelerating with large fraction of the cosmological density in
the form of a cosmological constant $\Lambda $. At a theoretical level $%
\Lambda $ is predicted to arise out of the zero-point quantum vacuum
fluctuations of the fundamental quantum fields. Using parameters arising in
the electroweak theory results in a value of the vacuum energy density $\rho
_{vac}=10^{6}$ GeV$^{4}$ which is almost $10^{53}$ times larger than the
current observational upper limit on $\Lambda $ which is $10^{-47}$ GeV$%
^{4}\sim 10^{-29}$ gm/cm$^{3}$. On the other hand the QCD vacuum is expected
to generate a cosmological constant of the order of $10^{-3}$ GeV$^{4}$
which is many orders of magnitude larger than the observed value. This is
known as the old cosmological constant problem. The new cosmological
constant problem is to understand why $\rho _{vac}$ is not only small but
also, as the current observations seem to indicate, is of the same order of
magnitude as the present mass density of the universe.

In what follows we are going to investigate how the value of the
cosmological constant changes in the context of the presence of the massless
quantum fields in the background of an Einstein universes having different
temperatures (or radii) as a consequence of the back reaction effect. This
investigation may be considered ''phenomenological'', in the sense that no
attempt is made to derive the model from an underlaying quantum field theory
in contrast to the customary approach normally adopted for such calculations.

Contracting the field equations in (\ref{q5}) we find that

\begin{equation}
\Lambda =\frac{R}{4}=\frac{3}{2a^{2}}.  \label{23}
\end{equation}

On the other hand the Einstein field equations reduces to

\begin{equation}
-\frac{3}{a^{2}}+\Lambda =-8\pi \rho _{tot},
\end{equation}

and

\begin{equation}
-\frac{1}{a^{2}}+\Lambda =\frac{8\pi \rho _{tot}}{3}
\end{equation}

where $\rho _{tot}=<T_{0}^{0}>_{tot}$. Solving the above two equations we
obtain

\begin{equation}
\Lambda =8\pi \rho _{tot}
\end{equation}

Using (\ref{23}) and the results obtained in the previous section for the
dependence of $T$ on $a$ we can solve for the dependence of $\Lambda $ on $T$%
. Fig. (3) depicts the relationship between the cosmological constant $%
\Lambda $ and the temperature for successive states of the Einstein universe
under the effect of back reaction of the neutrino field at finite
temperatures. It shows that the back reaction of finite-temperature quantum
fields in this model provide a large value for the cosmological constant
during the Casimir regime (vacuum dominated state of the universe). From the
point of view of inflationary models this large value of $\Lambda $ is
needed to resolve the problem of horizon and the problem of flatness, and
possibly to generate seed fluctuations for galaxy formation [19]. On the
other hand the value of the cosmological constant today, according to the a
above analysis, should be very small.

Fig. (4) compiles the data for the three massless fields: the scalar, the
neutrino and the photon field using the respective results for the radius
temperature relationship reported in I. It is clear that the contribution of
the scalar field to the cosmological constant is dominant, by an order of
magnitude, over all other fields throughout the Casimir regime and most of
the Planckian regime. However as we consider larger and larger spatial
sections of the Einstein universe the temperature goes asymptotically to
zero and the value of the cosmological constants for all the three fields
approaching one another.

By presenting these results we do not claim that the cosmological constant
problem is solved, but at least it may give us a glimpse of how the problem
is trivially solved within the context of an Einstein static universe.
Obviously the real universe is time-dependent, and a series of successive
instantaneously static states of the dynamical universe lacks all the
geometrodynamical effects; for example in this case particle production is
one important effect that is lost.

\section{Conclusions}

The present study exhibited some features of the thermodynamical behavior of
the Einstein static universe in presence of the neutrino field. In
presenting the results of this investigation we stress the fact that due to
the static nature of the Einstein universe, the following results are
specific to the case considered and should not be taken to imply an evolving
cosmological state. However, although not a realistic cosmological model,
the Einstein universe provide a useful theoretical model to achieve better
understanding of the interplay of spacetime curvature and of quantum field
theoretic effects. The main findings of this work reinforces the results
reported in I, with one additional point concerning the cosmological
constant, these are

(1) The thermal development of the universe is a direct consequence of the
state of its global curvature.

(2) The universe avoids the singularity at $T=0$ through quantum effects
(the Casimir effect) because of the non-zero value of $<T_{00}>_{0}$. A
non-zero expectation value of the vacuum energy density always implies a
symmetry breaking event.

(3) During the Casimir regime the universe is totally controlled by vacuum.
The energy content of the universe is a function of its radius. Using the
conformal relation between the static Einstein universe and the closed FRW
universe [10], this result indicates that in a FRW\ model there would be a
continuous creation of energy out of vacuum as long as the universe is
expanding, a result which was confirmed by Parker long ago [21]. The steep,
nearly vertical line in Fig. 2 suggests that the real universe started
violently and had to relax later.

(4) Throughout the Casimir regime the cosmological constant $\Lambda $ has a
large value for all three fields, and is nearly constant, but once the
universe goes to Planck regime the value of $\Lambda $ drops sharply. In the
realistic models for a dynamical universe, a large value for the
cosmological constant during the very early stages of the universe is
required so as to solve - via inflation- the horizon and flatness problem
[19]. On the other hand, the value for the cosmological constant at present
should be small in order to comply with observations. The results presented
in this paper, though may not be directly related to a developing universe,
provides us with the required behavior.

\begin{center}
\bigskip
\_\_\_\_\_\_\_\_\_\_\_\_\_\_\_\_\_\_\_\_\_\_\_\_\_\_\_\_\_\_\_\_\_\_\_\_\_\_%
\_\_\_\_\_\_

\textbf{References}
\end{center}

[\thinspace 1] Penzias A A and Wilson R W 1965 Astrophys. J \textbf{142} 419.

[2] Wilkinsoytler D T, O`Meara J M and Lubin D 2000 Physica Scripta \textbf{%
T85} 12.

[3] Ford L H 1975 Phys. Rev. D \textbf{11} 3370

[4] Dowker J S and Critchley R 1976 J. Phys. A\textbf{\ 9} 535

[5] Dowker J S and Altaie M B 1978 Phys. Rev. D \textbf{17} 417

[6] Gibbons G and Perry M J 1978 Proc. R. Soc. London, A \textbf{358} 467

[7] Bunch T S and Davies P C W 1977a Proc. R. Soc. London A \textbf{356} 569

-------------------------------------- 1977b Proc. R. Soc. London A \textbf{%
357} 381

-------------------------------------- 1978 Proc. R. Soc. London A \textbf{%
360} 117

[\thinspace 8] Birrell N D and Davies P C W (1982) \textit{Quantum Fields in
Curved Space} (Cambridge University Press,Cambridge, England ).

[9] Fulling S (1973) Phys. Rev. D \textbf{7} 2850

[10] Kennedy G (1978) J. Phys. A\textbf{11} L77

[11] Hu B L (1981), Phys. lett. B\textbf{103}, 331

[12] Plunien G Schutzhold R and Soff G (2000) Phys. Rev. Lett. \textbf{84}%
,1882

[13] Altaie M B (2002) Phys. Rev. D \textbf{65} 044028

[14] Schrodinger E. (1938), Comment. Pont. Acad. Sci. \textbf{2} 321

[15] Unruh W G (1974), Proc. R. Soc. London, A \textbf{338} 527

[16] Altaie M. B.(1978) and J. S. Dowker, Phys. Rev. D \textbf{18} 3557

[17] Kofman L A Sahni V and Starobinsky A A (1983) Sov. Phys. JETP \textbf{58%
} 1090

[18] Weinberg S (1972) \textit{Gravitation and Cosmology} (John Wiley \&
Sons, Inc., New York).

[19] Carroll S M (2000) \textit{The Cosmological Constant}, astro-ph/0004075
v2

[20] Sahni V and Starobinsky A A (2000), Int. J. Mod. Phys. D \textbf{9} 373

[21] Parker L (1969) Phys. Rev. \textbf{83} 1057

\bigskip

\textbf{Figure Caption}

FIG 1. The temperature-radius relationship for the neutrino field in an
Einstein universe.

FIG 2. Comparison beween the temperature-radius relationship for scalar,
neutrino and photon fields. The dashed line is for the scalar field, the
light solid line is for the neutrino field and the dark solid line is for
the photon field.

FIG 3. The variation of the cosmological constant with the temperature of
the universe resulting from the back reaction effect of the neutrino field.

FIG 4. Comparison between the contributions of the scalar, neutrino, and
photon fields to the cosmological constant in an Einstein universe at finite
temperatures. The dashed line is for the scalar field, the light solid line
is for the neutrino field and the dark solid line is for the photon field.

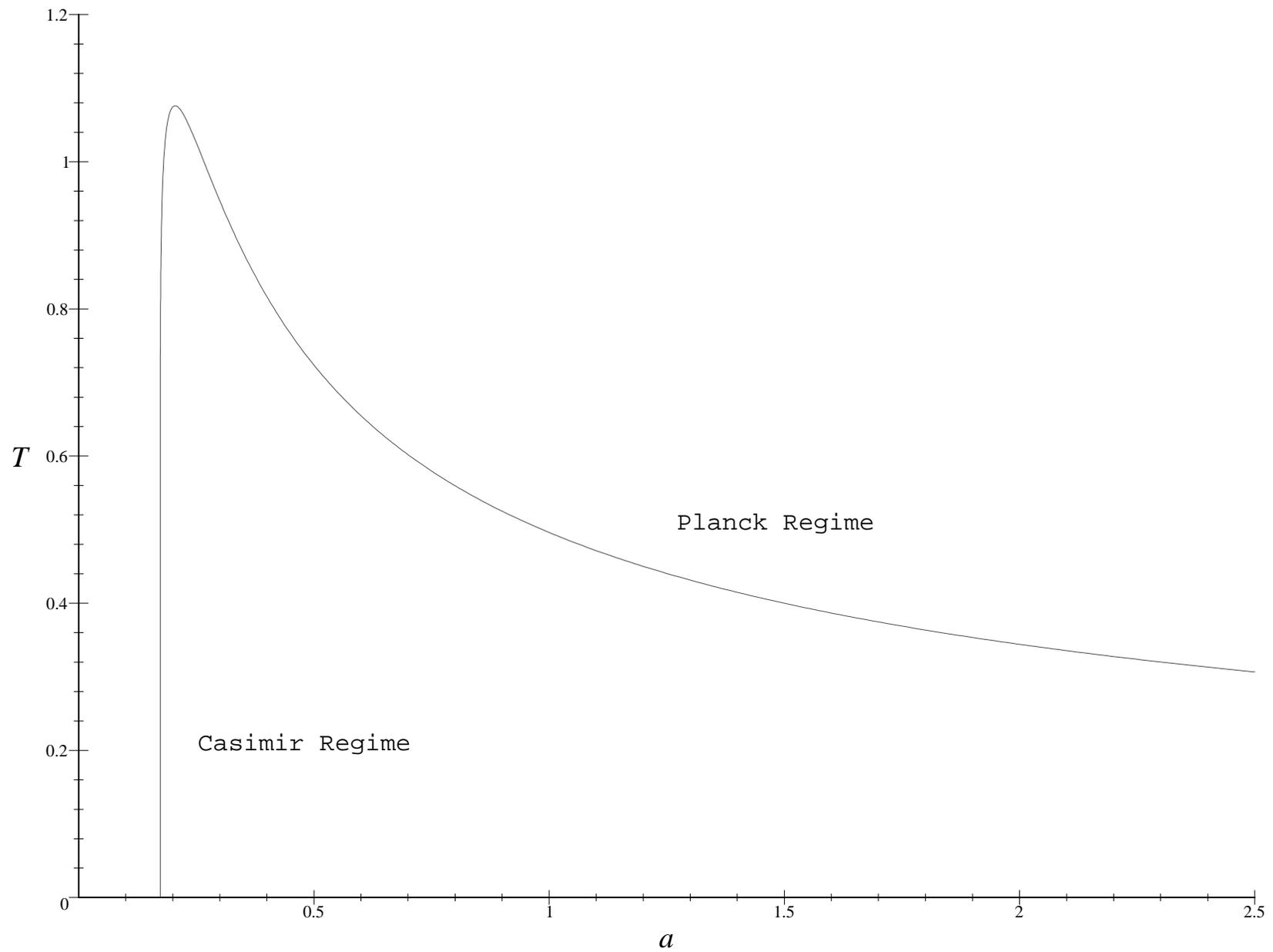
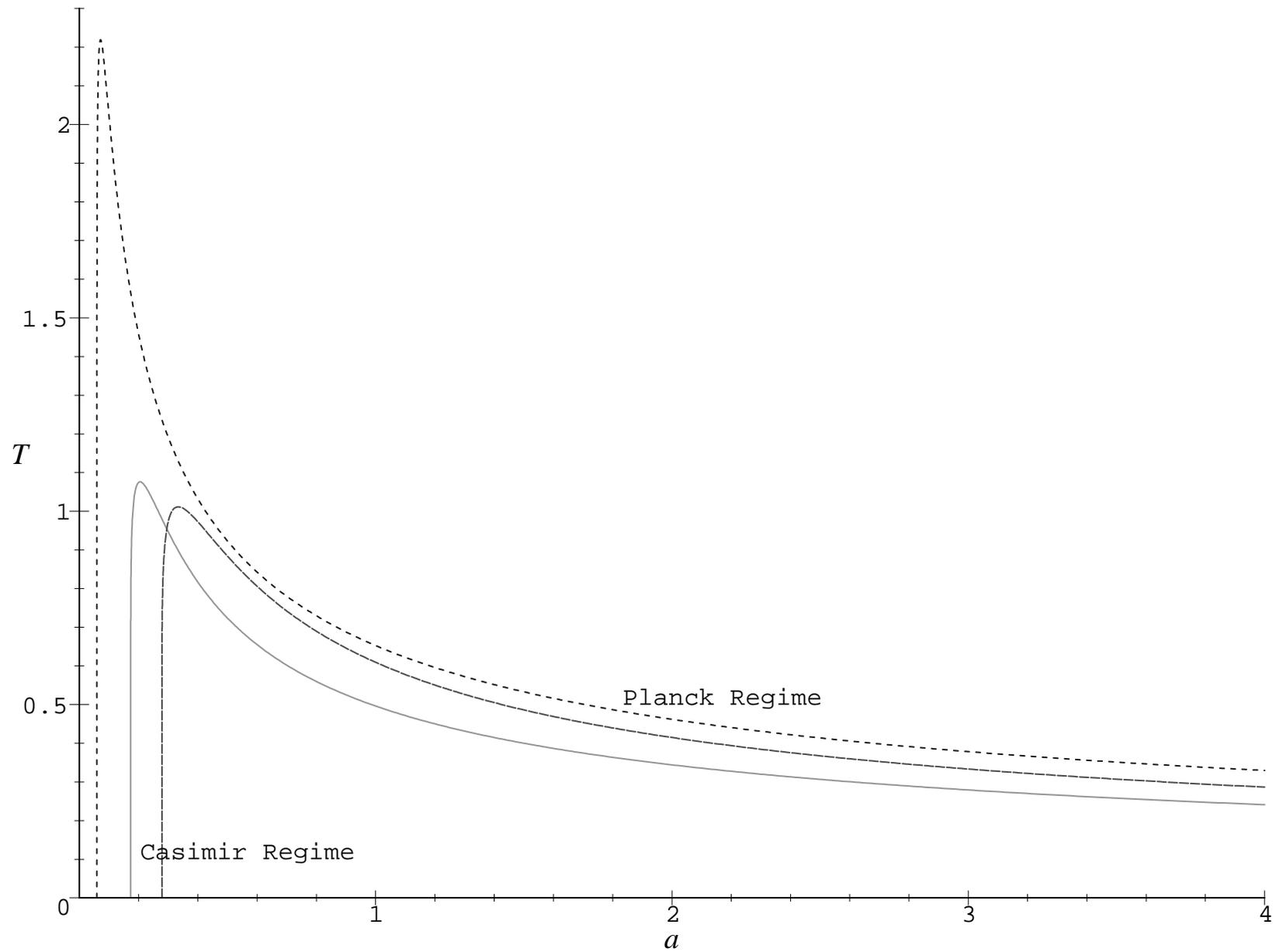
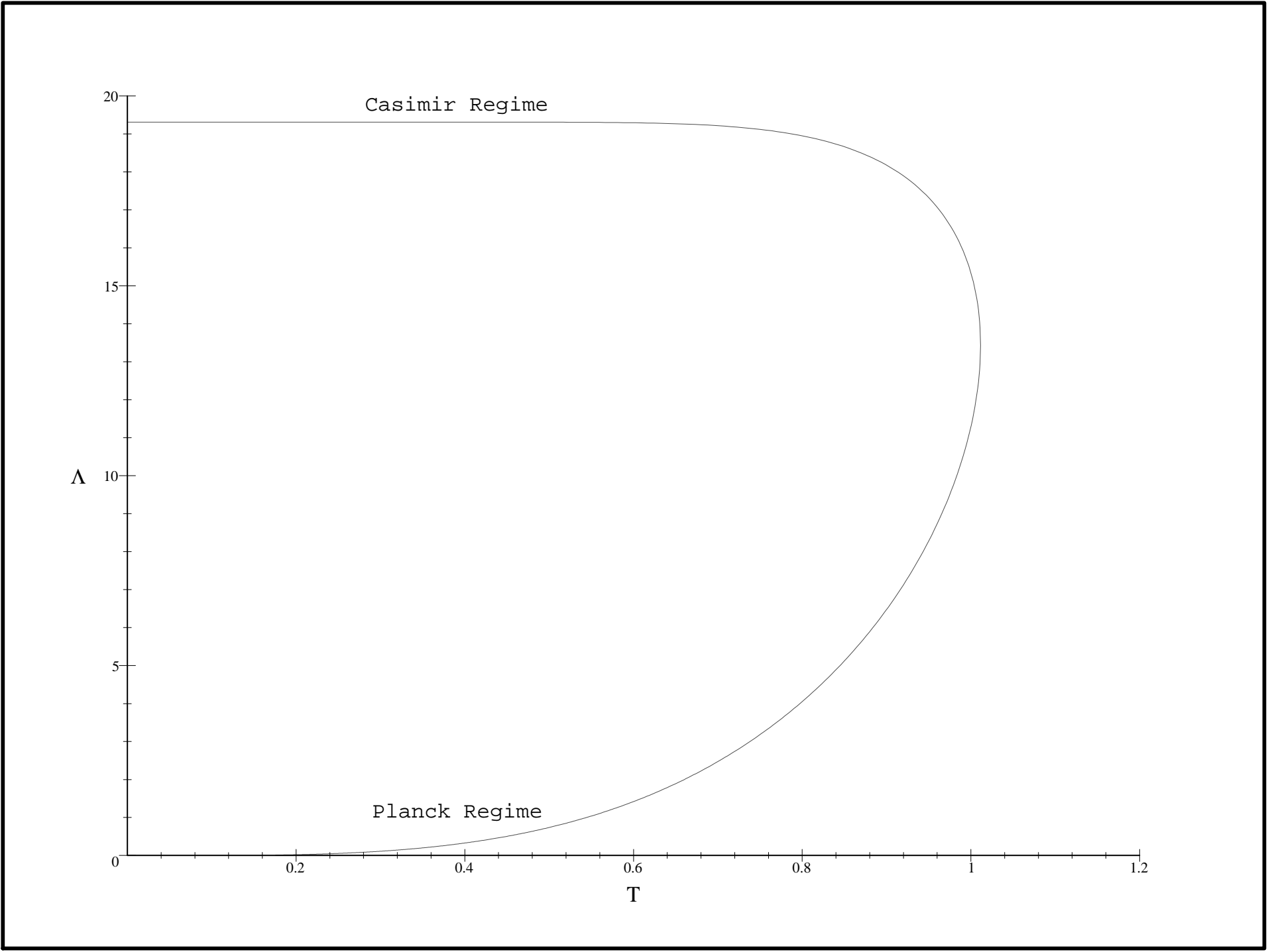
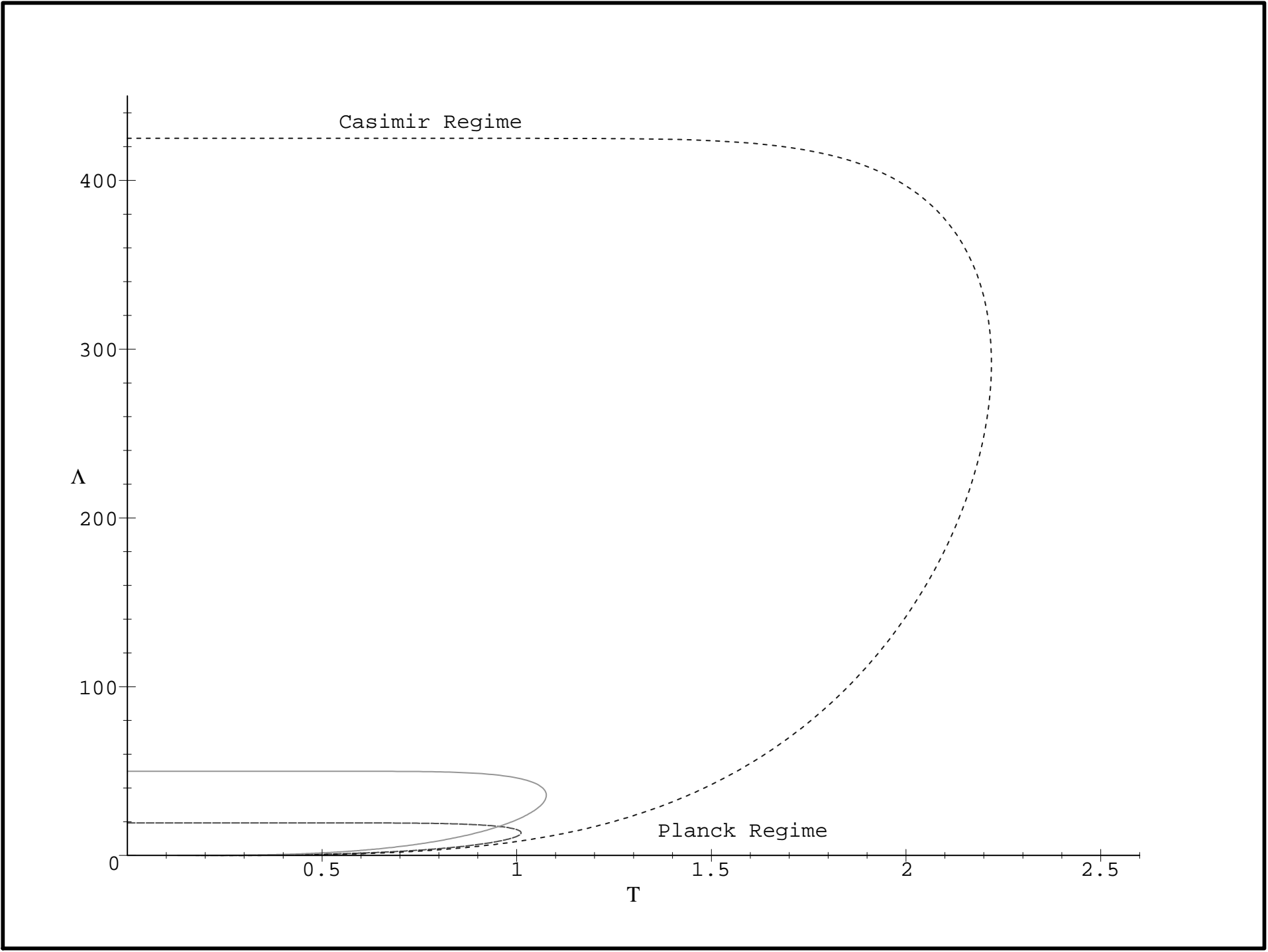
\end{document}